\documentclass[twocolumn,nopreprintnumbers,superscriptaddress,amsmath,pra]{revtex4}
\usepackage{graphicx}
\usepackage{amssymb}
\usepackage{array}

\begin{document}

\newcommand{\sqrtHz}{\ensuremath{\sqrt{\mbox{Hz}}}}
\newcommand{\microT}{\ensuremath{\mu\mbox{T}}}
\newcommand{\microHz}{\ensuremath{\mu\mbox{Hz}}}
\newcommand{\kOhm}{\ensuremath{\mbox{k}\Omega}}
\newcommand{\microVrms}{\ensuremath{\mu\mbox{V}_\mathrm{rms}}}
\newcommand{\microHzrms}{\ensuremath{\mu\mbox{Hz}_\mathrm{rms}}}
\newcommand{\mradrms}{\ensuremath{\mbox{mrad}_\mathrm{rms}}}

\newcommand{\taupower}[1]{\ensuremath{\tau^{#1}}}

\title{A sound card based multi-channel frequency measurement system}
\author{S.~Groeger}\affiliation{Physics Department, Universit\'e de Fribourg, Chemin du
Mus\'ee 3, 1700 Fribourg, Switzerland} \affiliation{Paul Scherrer
Institute, 5232 Villigen PSI, Switzerland}
\author{G.~Bison}\affiliation{Physics Department, Universit\'e de Fribourg, Chemin du
Mus\'ee 3, 1700 Fribourg, Switzerland}
\author{P.~E.~Knowles}\affiliation{Physics Department, Universit\'e de Fribourg, Chemin du
Mus\'ee 3, 1700 Fribourg, Switzerland}
\author{A.~Weis}\affiliation{Physics Department, Universit\'e de Fribourg, Chemin du
Mus\'ee 3, 1700 Fribourg, Switzerland}

\date{\today}

\begin{abstract}


For physical processes which express themselves as a frequency, for
example magnetic field measurements using optically-pumped
alkali-vapor magnetometers, the precise extraction of the frequency
from the noisy signal is a classical problem.  We describe herein a
frequency measurement system based on an inexpensive commercially
available computer sound card coupled with a software single-tone
estimator which reaches Cram\'er--Rao limited performance, a feature
which commercial frequency counters often lack.  Characterization of
the system and examples of its successful application to magnetometry
are presented.

\end{abstract}
\maketitle

\section{Introduction}

The present work is motivated by the need for a high resolution
frequency measurement system for analyzing signals generated by
optically-pumped cesium magnetometers~\cite{GroegerNIST}.  A set of
such magnetometers will be used for a detailed investigation of
magnetic field fluctuations and gradients in an experiment searching
for a neutron electric dipole moment (nEDM).  The experiment calls for
a magnetic field of between 1 to~2~\microT{} controlled at the 80~fT
level when measured over 100~s time intervals, control corresponding
to a relative uncertainty between 40 to 80~ppb.  The magnetometers are
based on the fact that for low magnetic fields the Larmor precession
frequency $f_L$ in a vapor of Cs atoms is proportional to the modulus
of the magnetic field $\vec{B}$
\begin{equation}
\label{eq:LarmorFreqtoField}
  f_L=\gamma  \left |\vec{B}\right|.
\end{equation}
The proportionality factor $\gamma$ is a combination of fundamental
and material constants and has a value of $\approx 3.5 \:
\mbox{kHz}/\microT$ for ${}^{133}\mbox{Cs}$.  The precession of the
atoms modulates the resonant absorption coefficient of the cesium
vapor, which is measured by a photodiode monitoring the power of a
laser beam traversing the atomic vapor~\cite{Blo62}. In the
self-oscillating mode of operation \cite{Blo62,CompLmpLas} the
magnetometer signal is of the form
\begin{equation}
    s(t)=A\sin{(2\pi f_L t+\phi)}+s_0\,.
\label{eq:magsignal}
\end{equation}
%
The Larmor frequency, $f_L$, has to be extracted from the signal.
Equation~\ref{eq:LarmorFreqtoField} connects the frequency
determination precision directly to the resulting field measurement
precision.  The basic demand on the frequency measurement system in
order to achieve the required field precision is a resolution of a few
hundred $\microHz$ in an integration time of 100~s.  Moreover, the
synchronous detection of signals from an array of magnetometers
requires a cost-effective multi-channel solution.

In our recent study \cite{CompLmpLas} of optically--pumped
magnetometer performance, frequency measurements were made with a
commercial frequency counter (Stanford Research Systems, model SR620),
which has a limited frequency resolution, thereby limiting the
magnetic field determination.  Frequency counters rely on the
detection of zero crossings of a periodic signal in a given dwell
time.  Their performance is limited by their resolution of the zero
crossing times, an event which is affected by the amplitude, offset,
and phase noise of the signal.  In demanding applications, such as
the one investigated here, that timing jitter limits the ultimate
frequency resolution of the magnetometer signal measurement.  Put
simple, the limitation of frequency counters is due to the fact that
they use only information in the vicinity of the zero crossings, while
valuable wave form information from in between the zero crossings is
ignored.

 As a more powerful alternative one can use numerical frequency
estimation algorithms to extract the frequency from the complete
waveform sampled at an appropriate rate and with a sufficient
resolution.  The performance of an ADC-based measurement system for
measuring a single frequency of about 8~Hz was discussed
in~\cite{Chibane1995}.  Under the assumption that a stable clock
triggers the ADC, the authors in \cite{Chibane1995} show that the
lower limit of the frequency resolution of their system coincides with
the Cram\'er--Rao lower bound (CRLB)~\cite{KayVol1}.  The CRLB is a
well-known concept from information theory and describes principle
limits for the estimation of parameters from sampled signals.

In our application, the Larmor frequency in a magnetic field of
$2\,\microT$ lies in the audio frequency range ($f_L=7\,\mbox{kHz}$).
We have investigated whether a commercially available (and rather
inexpensive) professional multi-channel sound card would present a
viable solution for sampling the magnetometer signals.  The estimation
of the frequency from the sampled data was done by a software
algorithm.  In the following we will show that such a simple system
can indeed be used for CRLB limited real-time frequency measurements
and for a detailed study of noise processes which limit the precision
of atomic magnetometers.

\section{The system}

The frequency measurement system consists of a professional sound card
(M--Audio Delta 1010) for digitizing the analog input data, an atomic
clock to provide~a stable time reference, and a standard personal
computer (PC) which reads the data and runs the frequency estimation
algorithm.  The sound card provides 8 analog input channels in a
breakout box that connects to a PCI interface card in the PC.  The
analog input signals can be sampled with a resolution of up to 24~bit
at a sampling rate of up to 96~kHz.  In order to limit the amount of
data we used only 16-bit resolution, which was proven to yield
sufficient precision.  Jitter or drifts of the sampling rate induce
additional phase noise on the sampled signal which can seriously
degrade the precision of the frequency estimation.  An essential
feature of the Delta 1010 sound card is its ``world clock'' input
which can be used to phase-lock the internal clock of the sound card
to an external 96~kHz time base.  The time base was realized by a
frequency generator synchronized to the 10~MHz signal of a rubidium
frequency standard (Stanford Research Systems, model PRS10).  The Rb
frequency standard provides a relative stability of $10^{-12}$ in
100~s which minimizes possible sampling rate jitter and drifts far
below the required level.  The requirements for the PC system are not
very demanding as long as it allows for the continuous recording of
the 16~bit data sampled at a rate of 96~kHz (5.8~GB/h for 8 channels).
A 1.8~GHz Pentium-4 processor was fast enough for real-time frequency
determination for all eight channels at a given integration time.
However, for the detailed analysis described below, in which the
integration time is varied, the time series were evaluated off-line
from the stored sampled data.

\section{Performance}

Considering the magnetometer signal given by Eq.~\ref{eq:magsignal},
the frequency $f_L$ is to be determined from the AC-coupled signal
data which, after sampling, are of the form
\begin{eqnarray}
x_n\! &\! =\! & \! \! A \sin{\left[
    2\pi\!\sum\limits_{k=1}^{n}
         \!\left(f_L + \delta f_k\right)\Delta t
        +\phi_0\!+\!\delta\phi_n
        \right]}\!+\delta x_n ,
\label{eq:sinussignal} \\
& &
 n = (0,\ldots,N-1) \nonumber \:,
\end{eqnarray}
where $A$ is the signal amplitude, $\Delta t$ the time resolution
(inverse of the sampling rate $r_s$), and $\phi_0$ the initial phase.
The number of sample points is $N=\tau/\Delta t\:(=\tau r_s)$, where
$\tau$ is the measurement integration time.
Also shown is the noise contribution at each point $n$ arising from
phase noise $\delta \phi_n$, frequency noise $\sum_{k=1}^{n} \delta
f_k$, and offset noise $\delta x_n$.
The frequency is determined from the data $x_n$ by a
maximum likelihood estimator based on a numerical Fourier
transformation which provides a CRLB limited value~\cite{CramerRao}.
The algorithm iteratively searches for the frequency $f$ that
maximizes the modulus of the Fourier sum
\begin{equation}
    MF(f)=\left| \sum_{n=0}^{N-1} x_n W_n
    \exp\left({i \frac{2 \pi f}{r_s}n}\right)
\right| \: ,
\label{eq:FourierSum}
\end{equation}
where $W_n$ is a windowing function.

Under ideal conditions (stable field and ideal electronics), the
frequency and phase noise ($\delta f_k$ and $\delta \phi_n$) are not
present in the signal (Eq.~\ref{eq:sinussignal}).  The fundamental
noise contribution is the photocurrent shot noise, which is
proportional to the square root of the DC offset $s_0$ in
Eq.~\ref{eq:magsignal}.  The noise is converted, by a transimpedance
amplifier, to voltage $V_{pc}$ and has a Gaussian amplitude
distribution with zero mean, corresponding to a white frequency
spectrum that is characterized by its power spectral density
$\rho_x^2$ (in $\mbox{V}^2/\mbox{Hz}$).  The signal-to-noise ratio
(SNR) is defined as $A^2/(\rho_x^2 f_s)$, where $f_s=(2 \Delta
t)^{-1}=r_s/2$ is the sampling rate limited bandwidth, i.e., the
Nyquist frequency or the highest frequency that can be detected
unaliased.  The CRLB of the frequency estimation from such an ideal
magnetometer signal is given by the variance~\cite{KayVol1}
\begin{equation}
\sigma^2_\mathrm{CRLB}=\frac{3 \rho_x^2}{\pi^2 A^2 \tau^3}.
\label{eq:CRfreq}
\end{equation}

\begin{table}[b]
\begin{tabular}{|>{$\vphantom{\sum\limits_j^k}$}l|r@{$\cdot$}c@{$\cdot$}l|l|} \hline
noise source    &\multicolumn{3}{c|}{$\sigma_A^2(\tau)$}     &
\multicolumn{1}{c|}{$\sigma_A(\tau)$}  \\ \hline\hline
white offset & 
$\displaystyle\frac{3}{\pi^2 A^2}$&$\displaystyle\rho_x^2$&
                 $\displaystyle\frac{1}{\tau^3}$ 
& $\propto\taupower{-3/2}$\\ \hline
flicker frequency&
       $\displaystyle 2\ln 2 $ & $\displaystyle h_f^{2}$ & $1$
& $\propto\taupower{0}$   \\ \hline
white frequency  &
$\displaystyle\frac{1}{2}$     & $\displaystyle\rho_f^2$ & $\displaystyle\frac{1}{\tau}$
& $\propto\taupower{-1/2}$\\ \hline
flicker phase    &
$\displaystyle\frac{3}{4\pi^2}$&$\displaystyle h_\phi^2$ &
                   $\displaystyle\frac{\ln(2\pi f_c \tau)}{\tau^2}$
& $\propto\taupower{-1}$  \\ \hline
white phase      &
$\displaystyle\frac{3}{8\pi}$  & $\displaystyle\rho_\phi^2$ & 
                   $\displaystyle\frac{f_c}{\tau^2}$
& $\propto\taupower{-3/2}$ \\ \hline
\end{tabular}
\caption{The central column shows the dependence of the Allan variance
$\sigma_A^2$ on the integration time $\tau$ and measurement bandwidth
$f_c$ for the noise sources listed at the left
\protect\cite{Audoin1979}.  White noise sources $\alpha$ are
characterized by their power spectral density $\rho^2_\alpha$. The
frequency dependent spectral density of flicker noise process $\alpha$
is modeled by $\rho^2_\alpha(f) = h^2_\alpha/f$.  By assuming the
relation $f_c=(2 \tau)^{-1}$ we find the power laws which typify
each noise type, shown in the right hand column.}
\label{table:bw-tau}
\end{table}

In frequency metrology it is customary to represent frequency
fluctuations in terms of the Allan standard deviation $\sigma_A$ ---
or its square $\sigma_A^2$, the Allan
variance~\cite{barnes,Audoin1979}.  One can show that for white noise
$\sigma_A$ coincides with the classical standard
deviation~\cite{Allan1987}.  A double logarithmic plot of the
dependence of $\sigma_A$ on the integration time $\tau$ is a valuable
tool for assigning the origin of the noise processes that limit the
performance of an oscillator (see for example
\cite{barnes,Audoin1979}).  As shown in Table~\ref{table:bw-tau}, the
variance $\sigma_A^2$ depends both on integration time $\tau$ and
measurement bandwidth $f_c$, which, for a measurement interval $\tau$,
is given by $f_c=(2 \tau)^{-1}$.  When that relation between bandwidth
and integration time is inserted into the formulas given in the
central column of Table~\ref{table:bw-tau} \cite{Audoin1979}, one
finds the typical $\tau$ dependencies of the Allan standard deviation
$\sigma_A$ shown in the right-hand column.  In the presence of several
uncorrelated noise processes, $\alpha$, the variance of the estimated
frequency is given by
\begin{equation}
\sigma^2 = \sum\limits_\alpha \sigma^2_\alpha(f)\:.
\label{eq:sigmaall}
\end{equation}
Note that for a magnetometer signal, the contribution from
Eq.~\ref{eq:CRfreq} will always be present in the sum.

\begin{figure}[t]
  \includegraphics[scale=1]{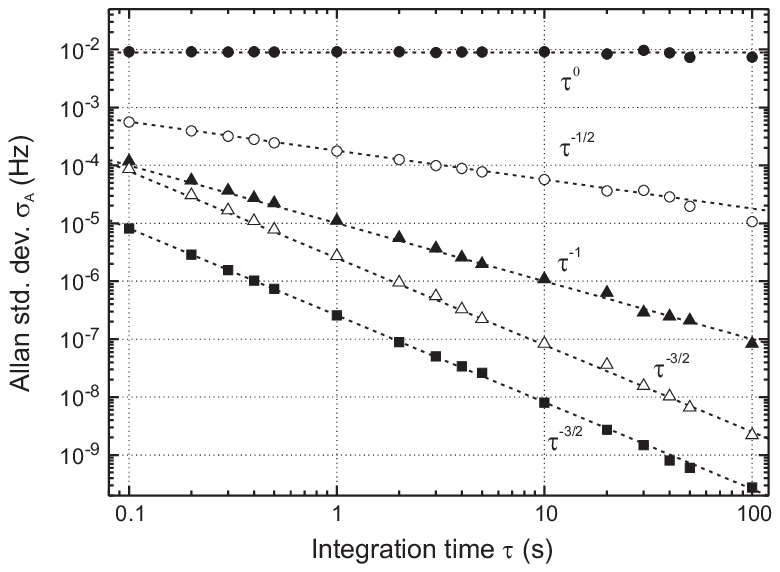}\\
  \caption{Allan standard deviation of the frequency of a synthesized
   sine wave affected by different noise processes. From top to
   bottom: flicker frequency noise (black dots), white frequency noise
   (open circles), flicker phase noise (black triangles), white phase
   noise (open triangles), white offset noise (black squares).}
\label{fig:simulation}
\end{figure}

We first investigated whether our data analysis algorithm reproduces
the theoretical $\tau$-dependencies shown in Table \ref{table:bw-tau}.
For that purpose we generated time series (16~bit, 96~kHz)
corresponding to Eq.~\ref{eq:sinussignal} with only one of the phase,
frequency, or offset noise terms enabled, and selected with well
defined spectral characteristics (flicker or white).
Figure~\ref{fig:simulation} shows the Allan standard deviation
$\sigma_A$ of those synthetic data. The emphasis here lies on the
slopes rather than on the absolute values, which were chosen to yield
a readable graph.

\begin{figure}[t]
  \includegraphics[scale=1]{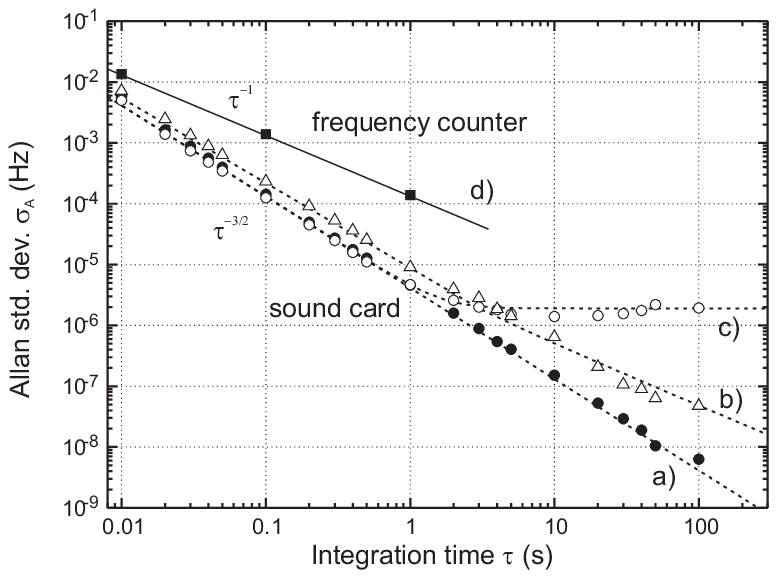}\\
  \caption{Allan standard deviation of the frequency of a sine wave
  affected by different noise processes, measured with the sound card
  (a--c) and a frequency counter (d): a) white offset noise, b) white
  offset noise and flicker phase noise, c) white offset noise and
  flicker frequency noise. The dashed lines represent the dependencies
  calculated on an absolute scale using the applied noise amplitudes.
  d) The same signal as in a) measured with a commercial frequency
  counter (Stanford Research model SR620) with a 300~Hz input bandpass
  filter.}
  \label{fig:FuncGenSolo}
\end{figure}

Next, we investigated the ability of the sound card to reach CRLB
limited detection of a 7~kHz sine wave.  The wave was generated by a
digital function generator (Agilent, model 33220A) stabilized to the
same Rb frequency standard as the sound card.  In order to simulate a
signal comparable to that of the magnetometers, the SNR of the
function generator output was artificially decreased from its nominal
value of better than $5\times 10^5$ to about $1.3\times 10^5$ (in a
1~Hz bandwidth) by adding white offset noise.
%
%
We recorded a 1~h time series of that signal, sampled with 16-bit
resolution.  The data were analyzed with the same algorithm as above
and yielded an Allan standard deviation $\sigma_A(\tau)$, shown as
black dots in Fig.~\ref{fig:FuncGenSolo}a).  The measurement agrees on
an absolute scale with the CRLB calculated using Eq.~\ref{eq:CRfreq}
and the applied SNR.  In addition to the offset noise, a second noise
source was used to apply $1/f$ noise, in turn, to the frequency or to
the phase modulation input of the function generator.  The resulting
$\sigma_A$ of the measured data is shown in
Figs.~\ref{fig:FuncGenSolo}b) and c).
Figure~\ref{fig:FuncGenSolo}d) shows $\sigma_A$ derived from the same
signal as Fig.~\ref{fig:FuncGenSolo}a) but analyzed by the commercial
frequency counter (Stanford Research Systems, model SR620) that was
used in~\cite{CompLmpLas}.  The three points shown correspond to the
three possible integration times of the SR620.  It can be clearly seen
that the counter technique does not allow the correct measurement of
these faint noise processes.  However, extrapolation of the data
points suggest that for integration times less than 10~ms the CRLB
could be reached.

\begin{figure}[t]
\includegraphics[scale=1]{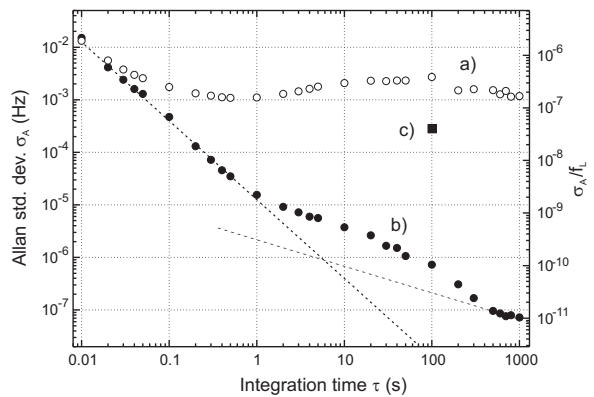}\\
\caption{a) ASD $\sigma_A$ of magnetic field fluctuations inside a
multi-layer magnetic shield.  b) Residual fluctuations of the
stabilized magnetic field.  The dashed lines indicate the CRLB
(left) and an assumed white frequency noise limitation (right).  c)
Stability required for the proposed nEDM experiment.}
\label{fig:R1stab}
\end{figure}

Finally, after the frequency estimator algorithm and the sound card
had proven their CRLB performance limit, we used the system to analyze
the frequency generated by an optically-pumped magnetometer (OPM).  A
magnetic field of $2\:\microT$ was produced by a solenoid driven by an
ultra-stable current source.  The OPM signal in that field is a 7~kHz
sine wave.  The OPM and the solenoid were located in a 6-layer
magnetic shield in order to suppress external field fluctuations.
Figure~\ref{fig:R1stab}a) shows $\sigma_A$ of a 2~h time series
recorded with the sound card.  The data represent pure magnetic field
fluctuations.  In particular, the approximately 2~mHz fluctuations in
the range between 2~to 200~s could be traced back to irregular current
fluctuations in the solenoid.  Nevertheless, the relative field
stability --- and therefore the relative current stability --- is on
the order of $3\times 10^{-7}$ for that range of integration times.
However, for a 100~s integration time the field instability exceeds
the requirement for the nEDM experiment mentioned in the introduction.

In order to determine the magnetometer performance limit, we actively
stabilized the magnetic field in the following way. The magnetometer
frequency was compared to a stable reference oscillator (i.e., the Rb
frequency standard) by means of a phase comparator, and the error
signal was used to control the solenoid current, thus realizing a
phase-locked loop.  Figure~\ref{fig:R1stab}b) shows the Allan standard
deviation of the OPM in the stabilized field, which is CRLB-limited up
to an integration time of 1~s.  The noise excess between 1~and 300~s
above the limits expected from the CRLB and the assumed white noise
limitation shows the limitation of the current stabilization scheme,
which nonetheless allows the suppression of the fluctuations by three
orders of magnitude at the integration time of interest.

We have realized a frequency measurement system based on a digital
sound card and have shown that it yields a performance superior to
commercial frequency counters.  We have proven that the system yields
CRLB limited frequency resolution in measurements of sine waves
affected by various sources of noise.  We have used the system to
prove that, at least in a limited range of integration times, an
active field stabilization by an optically pumped magnetometer is
limited by the theoretical Cram\'er-Rao bound.  The performance and
the multi-channel feature of the sound card and its external frequency
reference option present a low-cost alternative for applications
requiring simultaneous characterization of several frequency
generation systems, especially for long integration times.

\vfil

\section*{Acknowledgments}
We thank Francis Bourqui for help in developing the read-out
software.  We acknowledge financial support from Schweizerischer
Nationalfonds (grant no. 200020--103864), and Paul Scherrer Institute
(PSI).


\end{document}